\documentclass[jorunal]{IEEEtran}
\usepackage{subcaption}
\usepackage{mathrsfs}
\usepackage{caption}
\usepackage{amsmath}
\usepackage{epsfig}
\usepackage{bm}
\usepackage{slashbox}
\usepackage{algorithmic}
\usepackage{latexsym}
\usepackage{amsmath}
\usepackage{amssymb}
\usepackage{epsfig}
\usepackage{amsthm}
\usepackage{graphicx}
\usepackage[ruled,vlined]{algorithm2e}
\usepackage{epstopdf}

\newtheorem{proposition}{Proposition}

\begin{document}
%\begin{spacing}{1.9}

\title{{Multi-Dimensional Payment Plan in Fog Computing with Moral Hazard}}
\author{
Yanru Zhang,\thanks{Y. Zhang (yzhang82@uh.edu) and Z. Han (zhan2@uh.edu) are with the Department of Electrical and Computer Engineering, University of Houston, Houston, Texas 77004.}
Nguyen H. Tran,\thanks{N. H. Tran (nguyenth@khu.ac.kr) is with the Department of Computer Science and Engineering, Kyung Hee University, South Korea.}
Dusit Niyato,\thanks{D. Niyato (dniyato@ntu.edu.sg) is with the School of Computer Science and Engineering (SCSE), Nanyang Technological University, Singapore, 639798.}
and Zhu Han
}

%e-mail: hanzhu, kjrliu@glue.umd.edu}

%\date{}

%\setlength{\baselineskip}{25pt}
\maketitle
\pagenumbering{gobble}
\thispagestyle{empty}
\pagestyle{empty}
%\newpage
%\thispagestyle{empty}
% ------------------------------ ABSTRACT ---------------------------------
\begin{abstract}
Recently, the concept of fog computing which aims at providing time-sensitive data services has become popular. In this model, computation is performed at the edge of the network instead of sending vast amounts of data to the cloud. Thus, fog computing provides low latency, location awareness to end users, and improves quality-of-services (QoS). One key feature in this model is the designing of payment plan from network operator (NO) to fog nodes (FN) for the rental of their computing resources, such as computation capacity, spectrum, and transmission power. In this paper, we investigate the problem of how to design the efficient payment plan to maximize the NO's revenue while maintaining FN's incentive to cooperate through the moral hazard model in contract theory. We propose a multi-dimensional contract which considers the FNs' characteristics such as location, computation capacity, storage, transmission bandwidth, and etc. First, a contract which pays the FNs by evaluating the resources they have provided from multiple aspects is proposed. Then, the utility maximization problem of the NO is formulated. Furthermore, we use the numerical results to analyze the optimal payment plan, and compare the NO's utility under different payment plans.
\end{abstract}


%\setcounter{page}{1}
% ------------------------------ SECTION ---------------------------------
%%%%%%%%%%%%%%%%%%%%%%%%%%%%%%%%%%%%%%%%%%%%%%%%%%%%%%%%%%%%%%%
%%%%%%%%%%%%%%%%%%%%%%%%%%%%%%%%%%%%%%%%%%%%%%%%%%%%%%%%%%%%%%%
\section{Introduction}
%%%%%%%%%%%%%%%%%%%%%%%%%%%%%%%%%%%%%%%%%%%%%%%%%%%%%%%%%%%%%%%
%%%%%%%%%%%%%%%%%%%%%%%%%%%%%%%%%%%%%%%%%%%%%%%%%%%%%%%%%%%%%%%

The rapid developments of cloud computing have brought a centralized solution to application developers and content providers. Despite its wide known conveniences and advantages, cloud computing also suffers from certain limitations such as high latency and delay due to long distance between end users and servers \cite{Bonomi2012FCR}. The emerging trends in networking such as large distributed sensor networks, industrial automation, and high speed transportation that need location dependent fast processing and cannot be satisfied by the current service form by cloud computing \cite{Barbarossa2014SPM}.

With the motivation of placing the services as close as possible to end users, researchers have proposed a new cloud system called fog computing. In this model, fog nodes (FNs) such as end-user devices, access points, edge routers and switches are deployed at or very close to the edge of network, and with functionalities like converged computing, processing, management, networking, storage, physical and cyber security \cite{Patel2014}. Technically, fog computing is similar to cloud computing in the sense that both are made up of virtual systems providing the flexibility of on demand provisioning of compute, storage and network resources. However, fog computing has several advantages over cloud computing in the sense of a significant reduction in data movement across the network resulting in reduced congestion, cost and latency, elimination of bottlenecks resulting from centralized computing systems, improved security of encrypted data as it stays closer to the end user reducing exposure to hostile elements, and improved scalability arising from virtualized systems \cite{Cisco2015One}. By opening the access to fog computing nodes, service providers (SPs) can rapidly deploy certain applications and services to improve the quality of service (QoS) toward end users. This environment can also create a new value chain comprising NOs, InPs, SPs, and end users.

\begin{figure}[t]
    \centering
    \includegraphics[width=\columnwidth,height=0.22\textwidth]{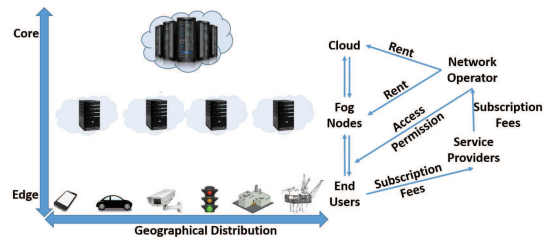}
    \caption{An illustration of fog computing system.}
    \label{fig:intro}
%    \vspace{-0.3cm}
\end{figure}
Referring to one brief model of fog computing in Fig. \ref{fig:intro}, there are variety of end users from the areas such as smart grid, industry, vehicular networks, transportation system, public safety department, and etc. that require real-time computing services. By subscribing to specific SPs whom subscribed to the network operator (NO) to obtain access to physical resources, end users are allowed to access computing resources in both the fog and cloud, with fog close to end users while cloud locates far away. Within this network, end users directly communicate with FN for real-time control and analytic, while the FNs only send periodic data summaries to the cloud for further aggregation and procession. Usually, the cloud and fog are managed by the NO, who rents the cloud center and fog nodes from infrastructure providers (InPs).

The NO is aiming at maximize its revenue by efficiently managing and coordinating the computing resources in both fog and cloud. To fully utilize the fog computing with minimal rent while ensuring the FNs receive non-negative revenue, appropriate payment plan is needed to enable the NO and FNs to play complementary roles within their respective business models, and allow all players to benefit from greater cooperation. Inspired by the effort-based reward from the labor market where employers pays its employees based on their work load, we propose a payment plan in fog computing such that the FNs receive their rental in accordance with the quantities of computing resources and the quality of service (QoS) they provide to the NO. The computing resources include the transmission bandwidth, power, computation capability (CPU speed), storage size, as well as a FN's proximity towards end users. Meanwhile the QoS can be referred to latency and delay during data transmission and processing, as well as security \cite{Sardellitti2015TSTPN}.

Based on this motivation, we aim at offering a contract that considers different aspects of the computing resources provided by FNs to end users, and assigns different payment weights in order to maximize the revenue of NO. Fortunately, the moral hazard model from contract theory provides us a useful tool to design such a payment plan that can solve the NO's revenue maximization problems in fog computing when the FN's performance in QoS is affected by multiple aspects \cite{Bolton2004}. From the NO's perspective, it ``employs'' the FNs to perform computing tasks and offers them QoS consistent payment by multi-dimension measurements. Inside this value chain, the NO tries to guarantee the fog computing QoS with minimal payment, while ensuring the FNs have necessary incentives to cooperate. Thus, to maximize the NO's revenue, the NO needs to find an optimal payment plan that can properly pay the infrastructure rent to FNs \cite{Werin.1992}.

The main contributions of this paper are summarized as follows. First, we are first to propose a QoS consistent contract which considers the quantities of resources provided by FNs from multiple aspects. The contract characterizes the general situation in real world and provides comprehensive payment plan to the FNs for using of those resources. Second, we formulate NO's revenue maximization problem, as well as provide the FNs with necessary incentive to cooperate in fog computing. Third, through simulations, we reveal different parameter's impacts on the optimal payment plan, and compare the NO utility under six different payment plans. Our results show that by using the proposed payment plan, the NO successfully maximizes the utilities and the FNs obtain the continuous incentives to participate in the fog computing.

The remainder of this paper is organized as follows. First, we will introduce the network model in Section \ref{sec:SystemModel}. Then, the problem formulation is described in Section \ref{sec:ProbForm}. The performance evaluation is conducted in Section \ref{sec:Simulations}. Finally, Section \ref{sec:Conclusion} draws the conclusion.

%\vspace{-0.4cm}
%%%%%%%%%%%%%%%%%%%%%%%%%%%%%%%%%%%%%%%%%%%%%%%%%%%%%%%%%%%%%%%
%%%%%%%%%%%%%%%%%%%%%%%%%%%%%%%%%%%%%%%%%%%%%%%%%%%%%%%%%%%%%%%
\section{System Model}\label{sec:SystemModel}

In this section, we consider a monopoly market with one NO trading with one FN. The NO-FN mutual benefit model is introduced first by constructing the multi-dimension payment plan offered by the NO. Then, we will give the utility functions of both the FN and NO before proceeding to the solution of the optimal contract. We assume that the NO considers $n$ aspects of the computing resources provided by FN and will pay the rent based on the QoS from the different aspects.
%\vspace{-0.4cm}
%%%%%%%%%%%%%%%%%%%%%%%%%%%%%%%%%%%%%%%%%%%%%%%%%%%%%%%%%%%%%%%
\subsection{Operation Cost}\label{subsec:cost}
%%%%%%%%%%%%%%%%%%%%%%%%%%%%%%%%%%%%%%%%%%%%%%%%%%%%%%%%%%%%%%%
%\vspace{-0.3cm}

\begin{figure}[t]
    \centering
    \includegraphics[width=1\columnwidth,height=0.27\textwidth]{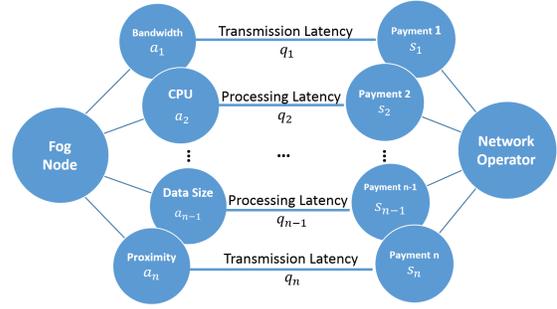}
    \caption{The multi-dimension resource quantity and QoS consistent payment contract.}
    \label{fig:contract}
%    \vspace{-0.3cm}
\end{figure}

In fog computing, the FN encounters both capital expenditure (CapEx) and operational expenditure (OpEx) to provide heterogeneous resources. CapEx is the prominent investment which includes the cost of purchasing and installing equipments such as routers, switches, access points, backhaul aggregators, as well as the cost of using licensed spectrum issued by the authorities \cite{Cisco2015Two}. Meanwhile the OpEx includes the energy consumption due to signal processing, execution, and data transmission. We assume that the CapEx is fixed, while the OpEx is usage based.

The FN's heterogeneous resources are often measured in disparate scales or units. It has been proved that those measurements can be mapped into one single unit, such as time \cite{Nishio2013SHR}. Thus by mapping and normalization, we can represent a FN's contribution of resources such as bandwidth, CPU, transmission power by a vector $a=(a_1,\ldots, a_n)$, $n \geq 1$ for one computing task. After such mapping, each $a_i$ has the same scale or unit and represents one resource type. Such mapping is based on the knowledge that the bandwidth and CPU speed affect the transmission and processing time, respectively. Due to the context aware and location dependent properties of fog computing, the size of data being processed and the geographic distance between FN and end users also have the impacts on data processing and transmission latency. There are many other aspects that affect the QoS of fog computing that we have not listed in Fig. \ref{fig:contract}, such as transmission power, which can also be mapped to transmission latency.

When providing those resources, the FN's cost incurred is defined in a quadratic form,
%\vspace{-0.3cm}
\begin{equation}
\psi(a)=\frac{1}{2} a^T C a,
%\vspace{-0.3cm}
\end{equation}
where $C$ is a symmetric $n\times n$ matrix with the form of
\begin{equation}
C=
\begin{bmatrix}
   c_{11}      & \cdots & c_{1n}
\\ \vdots & \ddots & \vdots
\\ c_{n1}      & \cdots & c_{nn}
\end{bmatrix}.
%\vspace{-0.3cm}
\end{equation}
The diagonal element $c_{ii}$ of $C$ reflects the FN¡¯s resource-specific cost coefficient, and the off-diagonal elements $c_{ij}$ represent the cost relationship between two resources $i$ and $j$.

The sign of $c_{ij}$ indicates technologically substitute, complementary, independent between two resources $i$ and $j$, if $c_{ij} > 0$, $< 0$, $= 0$, respectively. If two resources are technologically substitute, raising the quantity of one resource raises the marginal cost of the effort on the other resources. The example of technologically substitute is the relationship between geographic distance and transmission power. To achieve the same data rate at the end user, longer distance requires a higher transmission power consumption. In contrast, raising the quantity on one resource decreases the marginal cost of the other resource if they are technologically complementary. One example is about the relationship between bandwidth and transmission power. Given the same data package size and transmission distance, the larger bandwidth can achieve the same data rate at receiver with lower transmission power. In this example, high quality in one resource ease the cost in the other, and thus called technologically complementary. For technologically independent resources, their operation cost is not dependent on the quantity of other resources. There are many technologically independent examples in fog computing, such as the relationships between transmission bandwidth and CPU speed, geographic distance and data size.

In order to lower the mathematical complexity, we only solve the cases without the technologically complementary in this paper. Thus, the operation cost coefficient matrix is a positive semi-definite matrix with every element in $C$ is non-negative.

%\vspace{-0.5cm}
%%%%%%%%%%%%%%%%%%%%%%%%%%%%%%%%%%%%%%%%%%%%%%%%%%%%%%%%%%%%%%%
\subsection{QoS Measurement}\label{subsec:measure}
%%%%%%%%%%%%%%%%%%%%%%%%%%%%%%%%%%%%%%%%%%%%%%%%%%%%%%%%%%%%%%%
%\vspace{-0.3cm}
The resources such as CPU speed, transmission bandwidth and power can be easily specified by the FN, the end user related parameters such as geographic distance, data size, can also be quantified easily, while the measurement of QoS cannot be that accurate. Though those measurements can be mapped into time scale follow the procedure in \cite{Nishio2013SHR}, error may come from the design failure of the measurement system.

Given the actual resources provided by FN is $a$, which is hidden from the NO, but the FN's QoS can be observed as a vector of QoS $q=(q_1,\ldots, q_n)$, $n \geq 1$, which can be regarded as the FN's performance in latency reduction. Due to aforementioned different measurability on QoS, the received information $q$ varies. Therefore, the performance of the FN is a noisy signal of the resources it has provided:
%\vspace{-0.35cm}
\begin{equation}
q=a+\varepsilon,
%\vspace{-0.35cm}
\end{equation}
where the random component $\varepsilon=(\varepsilon_1,\ldots, \varepsilon_n)$, $n \geq 1$, is assumed to be normally distributed with mean zero and covariance matrix $\Sigma$. Thus, the FN's performance follows the distribution of $q\sim N(a,\Sigma)$.

The variance $\Sigma$ is a symmetric $n\times n$ covariance matrix with the form of
\begin{equation}
\Sigma=
\begin{bmatrix}
   \sigma_1^2      & \cdots & \sigma_{1n}
\\ \vdots & \ddots & \vdots
\\ \sigma_{n1}      & \cdots & \sigma_n^2
\end{bmatrix},
%\vspace{-0.2cm}
\end{equation}
where $\sigma_i^2$ denotes the variance of $\varepsilon_i$, and $\sigma_{ij}$ is the covariance of $\varepsilon_i$ and $\varepsilon_j$. The variance denotes the difficulty to guarantee the correctness of measuring the QoS, and also reflects the resource quantity difference observed at the FN and NO sides. If the variance is large, the measurability of the QoS is difficult, and there is a high probability that the FN's performance is poorly measured and far away from the true amount of resource the FN has provided. If the QoS is easy to measure, the variance will be small or even zero. For example, the data size is an independent measure with variance $0$, as well as the geographic distance. While the data processing time not only depends on the data size, but also the complexity of algorithm, which has a large variance. The covariance of two measurements exists because the measurement of one resource may affect the measurement of the others; for example the transmission time is affected by both bandwidth and power.

%\vspace{-0.4cm}
%%%%%%%%%%%%%%%%%%%%%%%%%%%%%%%%%%%%%%%%%%%%%%%%%%%%%%%%%%%%%%%
\subsection{Payment Plan}\label{subsec:reward}
%%%%%%%%%%%%%%%%%%%%%%%%%%%%%%%%%%%%%%%%%%%%%%%%%%%%%%%%%%%%%%%
%\vspace{-0.2cm}
Inspired by the manager's reward package in industry, which comprises a fixed salary, a bonus related to the firm's profits, and stock options related reward based on the firm's share price \cite{Bebchuk2002}, we define the FN's payment plan $w$ in fog computing as a linear combination of a fixed salary and QoS related payments. By restricting the payment plan offered by the NO in the linear form, the payment plan $w$ FN receives by participating in the fog computing can be written as
%\vspace{-0.35cm}
\begin{equation}
w=t+s^Tq,
%\vspace{-0.3cm}
\end{equation}
where $t$ denotes the fixed salary, which is a constant independent of QoS and regarded as the subscribing fee to complement the FN's CapEx. $s=(s_1,\ldots, s_n)$, $n \geq 1$ is the payment related to the QoS $q$. As $q$ is a random variable which follows $q\sim N(a,\Sigma)$, the payment plan $w$ is also a random variable with a mean of $t+s^Ta$. From the scaling property of covariance, we know that $Var(s^Tq)=s^T \Sigma s$. Thus, the payment plan follows the distribution $w\sim N(t+s^Ta,s^T \Sigma s)$.

At this point, we can propose the contract that is offered by the NO as ($a,t,s$), where $a$ and $s$ are $n\times 1$ vectors, and $t$ is a constant value. Under this contract, the NO offers the FN a payment plan which includes a fixed salary $t$, and $n$ QoS related payments $(s_1,\ldots, s_n)$. Fig. \ref{fig:contract} illustrates how this contract works. The FN provides the quantity $a_i$ for resource $i$ in the computing task, which is observed as a QoS $q_i$ by the NO. The NO then offers a payment $s_i$ related to $q_i$.

%\vspace{-0.4cm}
%%%%%%%%%%%%%%%%%%%%%%%%%%%%%%%%%%%%%%%%%%%%%%%%%%%%%%%%%%%%%%%
\subsection{Utility of Fog Node}\label{subsec:FN}
%%%%%%%%%%%%%%%%%%%%%%%%%%%%%%%%%%%%%%%%%%%%%%%%%%%%%%%%%%%%%%%
%\vspace{-0.3cm}
In this model, we assume that the FN has constant absolute risk averse (CARA) risk preferences, which means the FN has a constant attitude towards risk as its income increases. Such a risk preference comes from the FN's concern about its security issue when opening access for end users. Thus, FN utility is represented by a negative exponential utility form \cite{Norstad1999},
%\vspace{-0.3cm}
\begin{equation}\label{eq:Uutility}
u(a,t,s)=-e^{-\eta[w-\psi(a)]},
%\vspace{-0.3cm}
\end{equation}
where $\eta>0$ is the FN's degree of absolute risk aversion
%\vspace{-0.3cm}
\begin{equation}
\eta = -\frac{u''}{u'},
%\vspace{-0.3cm}
\end{equation}
where $u$ is the FN's utility function. A larger value of $\eta$ means more incentive for the FN to provide more resources for the computing task. The utility and operation cost of the FN are measured in such monetary units that they are consistent with the payment from the NO.

From (\ref{eq:Uutility}), we see that the FN's utility is a strictly increasing and concave function. For lower computation complexity, we can make use of the exponential form of the utility function, and use \emph{certainty equivalent} as a monotonic transformation of the FN's expected exponential utility function \cite{CE2003}.
\begin{proposition}
The FN's utility can be equally represented by certainty equivalent:
\begin{align}
CE_u =t+s^Ta-\frac{1}{2} a^T C a-\frac{1}{2}\eta s^T\Sigma s.
\end{align}
\end{proposition}
The certainty equivalent consists of the expected payment minus the operation cost and measurement cost. %The detail proof of this transformation can be found in \ref{}.

%\vspace{-0.4cm}
%%%%%%%%%%%%%%%%%%%%%%%%%%%%%%%%%%%%%%%%%%%%%%%%%%%%%%%%%%%%%%%
\subsection{Utility of Network Operator}\label{subsec:princ}
%%%%%%%%%%%%%%%%%%%%%%%%%%%%%%%%%%%%%%%%%%%%%%%%%%%%%%%%%%%%%%%
%\vspace{-0.3cm}
Here, we define the utility of the NO as the expected gross benefits of $V(a)$ minus the payment plan $w$ to the FN. Thus, the NO's expected utility is written as
%\vspace{-0.3cm}
\begin{equation} %\vspace{-0.1cm}
U(a,t,s)=V(a)-w,
%\vspace{-0.3cm}
\end{equation}
where $V(\cdot)$ is the evaluation function which follows $V(0)=0$ and $V'(\cdot)>0$. Different from the FN who has CARA risk preferences, the NO here is assumed to be risk neutral, i.e., $V''(\cdot)= 0$. Thus, the expected profit of the NO can be simplified to
%\vspace{-0.3cm}
\begin{align}% \vspace{-0.1cm}
U(a,t,s)= \beta^Ta-w,
%\vspace{-0.3cm}
\end{align}
where $\beta=(\beta_1,\ldots, \beta_n)$, $n \geq 1$, characterizes the marginal effect of the FN's contribution $a$ on the NO's utility $V(a)$. Similar to the definition of FN's certainty equivalent, we can derive the NO's certainty equivalent as
%\vspace{-0.3cm}
\begin{align}% \vspace{-0.1cm}
CE_p&= E[\beta^Ta-w],\\ \nonumber
    &= \beta^Ta-s^Ta-t.
%\vspace{-0.8cm}
\end{align}

%\vspace{-0.8cm}
%%%%%%%%%%%%%%%%%%%%%%%%%%%%%%%%%%%%%%%%%%%%%%%%%%%%%%%%%%%%%%%
\subsection{Social Welfare}\label{subsec:social}
%%%%%%%%%%%%%%%%%%%%%%%%%%%%%%%%%%%%%%%%%%%%%%%%%%%%%%%%%%%%%%%
%\vspace{-0.3cm}

With the definitions of both FN's and NO's utility functions and certainty equivalent payoffs, we can have the social welfare defined as their joint surplus, i.e., the summation of FN's and NO's equivalent certainty:
%\vspace{-0.3cm}
\begin{align}
R&=CE_u+CE_p, \\ \nonumber
 &=\beta^Ta-\frac{1}{2} a^T C a-\frac{1}{2}\eta s^T\Sigma s.
% \vspace{-0.3cm}
\end{align}
The social welfare is the resource provided by the FN, minus the operation cost and the cost incurred by inaccurate measurement. Notice that this expression is independent of the fixed salary $t$, which serves as an intercept term in the contract. Thus, the fixed salary $t$ can only be used to allocate the total certainty equivalent between the two parties \cite{Holmstrom1991JLEO}.

\begin{figure*}
 \begin{subfigure}[b]{0.32\textwidth}
 \centering
 \includegraphics[width=\columnwidth,height=0.81\textwidth]{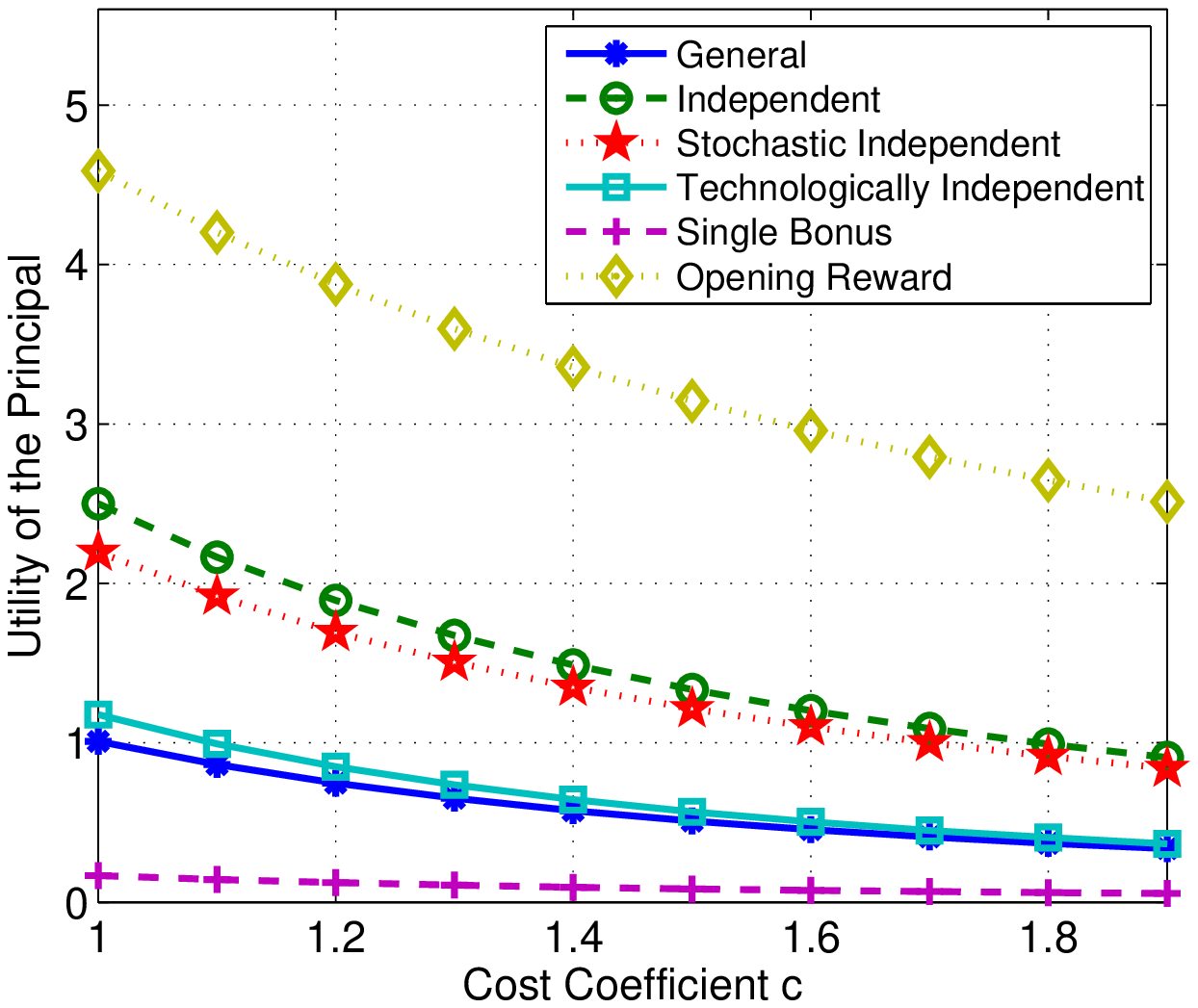}
 \caption{Cost coefficient $c_{ii}$}
 \label{fig:costP}
 \end{subfigure}
  \begin{subfigure}[b]{0.32\textwidth}
 \centering
 \includegraphics[width=\columnwidth,height=0.81\textwidth]{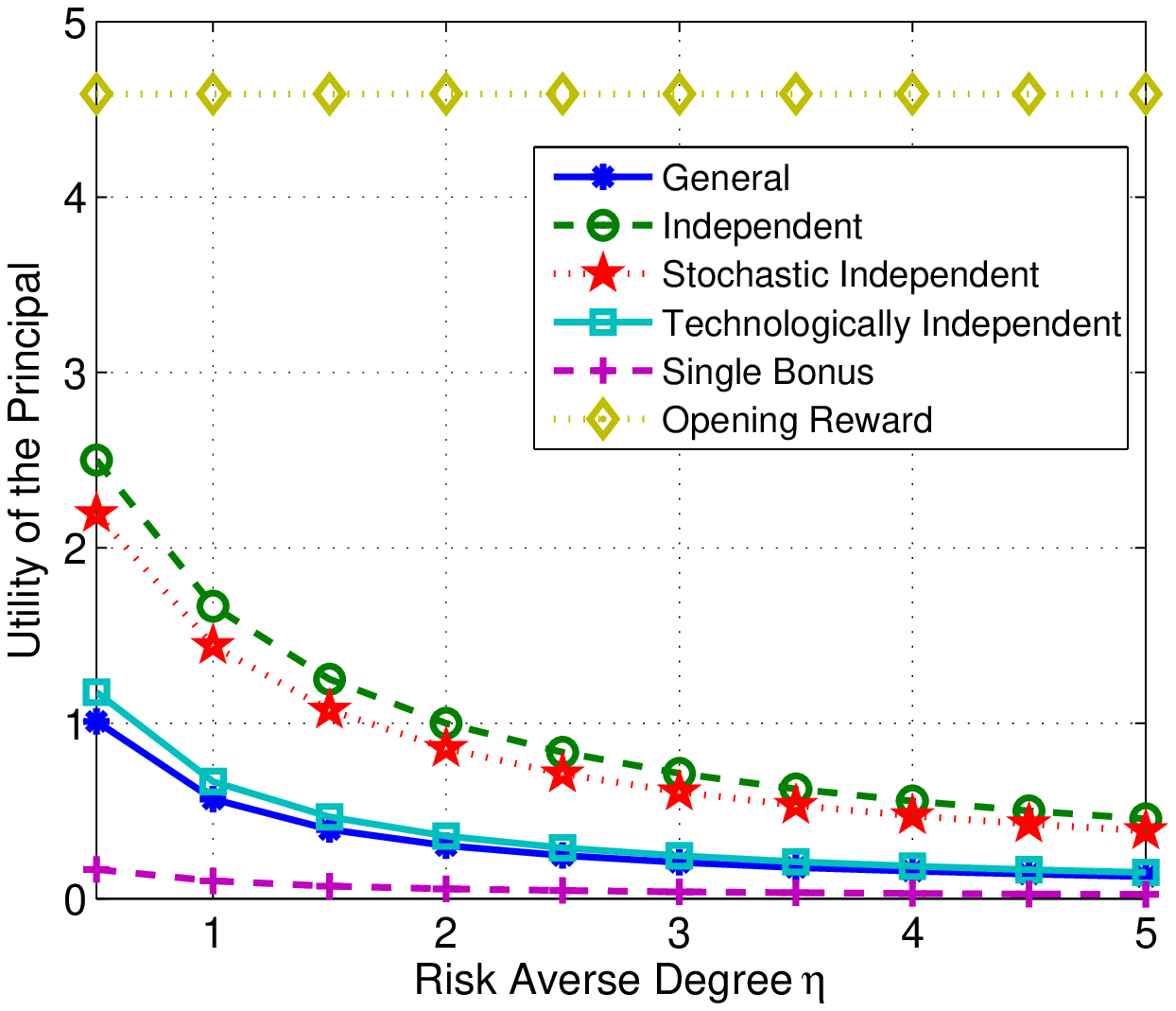}
 \caption{Risk averse degree $\eta$}
 \label{fig:riskdgreeP}
 \end{subfigure}
   \begin{subfigure}[b]{0.32\textwidth}
 \centering
 \includegraphics[width=\columnwidth,height=0.81\textwidth]{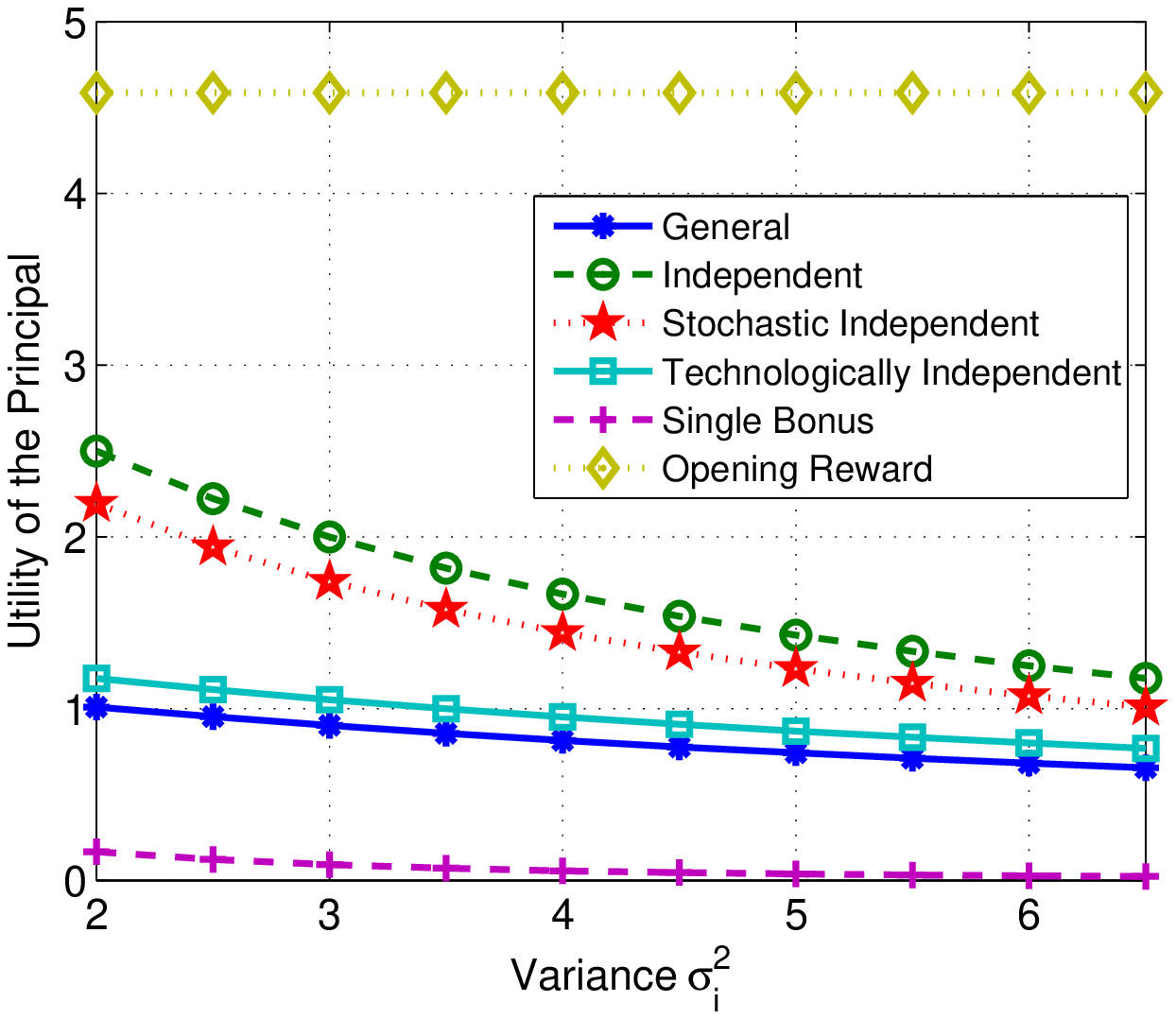}
 \caption{Measurement error variance $\sigma_{i}^2$}
 \label{fig:varianceP}
 \end{subfigure}
\caption{The NO's utility as different parameters vary}
\label{fig:NO}
\vspace{-0.3cm}
\end{figure*}

%\vspace{-0.4cm}
%%%%%%%%%%%%%%%%%%%%%%%%%%%%%%%%%%%%%%%%%%%%%%%%%%%%%%%%%%%%%%%
%%%%%%%%%%%%%%%%%%%%%%%%%%%%%%%%%%%%%%%%%%%%%%%%%%%%%%%%%%%%%%%
\section{Problem Formulation}\label{sec:ProbForm}
%%%%%%%%%%%%%%%%%%%%%%%%%%%%%%%%%%%%%%%%%%%%%%%%%%%%%%%%%%%%%%%
%%%%%%%%%%%%%%%%%%%%%%%%%%%%%%%%%%%%%%%%%%%%%%%%%%%%%%%%%%%%%%%
%\vspace{-0.3cm}
With the system model, we can formulate the NO's utility maximization problem while providing the FN necessary incentives to cooperate. The NO's problem can be written as
%\vspace{-0.3cm}
\begin{eqnarray}\label{eq:Opt1}
          && \max_{a,t,s} \quad U(a^*,t,s),\\
s.t.      &&(a) \quad\!\! a^* \in \arg\max_a   u(a,t,s), \nonumber \\
          &&(b) \quad\!\! u(a^*,t,s) \geq u(\overline{w}), \nonumber
%\vspace{-0.3cm}
\end{eqnarray}
where $u(\overline{w})$ is the reservation utility of the FN when not providing any resource ($a=\textbf{0}$) in the fog computing. The NO maximize its own utility under the incentive compatible (IC) constraint (a) that the FN provides the optimal amount of resource $a^*$ maximizing its own utility, and the individual rationality (IR) constraint (b) that the utility FN received is no less than its reservation utility.

Under the assumption of stochastic dependent, the error terms are stochastically interacted, i.e., $\sigma_{ij} \neq 0$. For technologically dependent, we mean that the activities are technologically correlated with each other, i.e., $c_{ij}> 0$ and $C$ is a positive definite matrix. We solve this multi-dimensional problem by using certainty equivalent model with the following simple reformulation of the NO's problem:
%\vspace{-0.3cm}
\begin{eqnarray}\label{eq:Opt2}
     &&\max_{a,t,s} \quad \beta^Ta- s^Ta - t,\\
s.t. && (a) \quad\!\! a^* \in \arg\max_a [t + s^Ta- \frac{1}{2} a^T C a -\frac{1}{2}\eta s^T\Sigma s],  \nonumber
\\
     && (b) \quad\!\!t + s^Ta -\frac{1}{2} a^T C a -\frac{1}{2}\eta s^T\Sigma s \geq \overline{w}, \nonumber
%\vspace{-0.3cm}
\end{eqnarray}
where $\overline{w}$ also denotes the reservation utility of the FN when not participating in the fog computing. The IC constraint represents the rationality of the FN¡¯s choice of contribution. The IR constraint in (b) ensures that the NO cannot force the FN into accepting the contract.

We first solve the optimal effort by reducing the IC constraint first. The FN's certainty equivalent is concave, since its second-order derivative with respect to $a$ is a negative definite matrix $- C$. Thus, the optimal effort can be determined by taking the first-order derivative of the FN's certainty equivalent regarding $a$, and set $u'(a,t,s)=0$. In the matrix differentiation, if we define $\alpha=a^T C a$, as $C$ is a symmetric matrix, we have $\partial \alpha/\partial a=2 a^T C$ \cite{Barzel1982}. Since $C$ is symmetric positive definite, its inverse is existent. Thus, through numerical derivations, we finally have $a=C^{-1}s$ in this multi-dimension case. Accordingly, we substitute the IR constraint in (b) with the optimal amount of resource $a^*$ and simplify the NO's problem to
%\vspace{-0.3cm}
\begin{eqnarray}\label{eq:Opt3}
&& \max_{a,t,s} \quad \beta^T C^{-1}s -s^T C^{-1}s- t,
\\
s.t. && (a) \quad\!\!  t+s^{T} C^{-1}s- \frac{1}{2} (C^{-1}s)^{T} C (C^{-1}s) \nonumber
\\   && \quad\quad\quad\quad\quad\quad\quad\quad\quad\quad\quad\quad\quad\quad  -\frac{1}{2}\eta s^{T}\Sigma s=\overline{w}. \nonumber
%\vspace{-0.4cm}
\end{eqnarray}
Substituting the value of $t$ in the IR constraint to the objective and differentiating the objective function with respect to $s$, we have the QoS related payment $s^*$ in the optimal multi-dimension payment plan as:
%\vspace{-0.5cm}
\begin{align}\label{eq:effort}
s^* =(C^{-1} + \eta \Sigma)^{-1} C^{-1}\beta=  (I + \eta C\Sigma)^{-1}\beta.
%\vspace{-0.5cm}
\end{align}
With $s^*$, we have the optimal amount of computing resource in the multi resource case as
%\vspace{-0.4cm}
\begin{align}
a^*=C^{-1}(I + \eta C\Sigma)^{-1}\beta.
%\vspace{-0.5cm}
\end{align}
Representing $t$ by $\overline{w}$, $s^*$ and $a^*$, we obtain the fixed payment $t$ in the optimal linear payment plan as:
%\vspace{-0.4cm}
\begin{align}
&t^* = \overline{w}+\frac{1}{2}s^{T}(\eta \Sigma-C^{-1})s, \\  \nonumber
    & = \overline{w}+\frac{1}{2} \left[(I + \eta C\Sigma)^{-1}\beta\right]^{T} (\eta \Sigma -C^{-1}) \left[ (I + \eta C\Sigma)^{-1}\beta \right].
%\vspace{-0.5cm}
\end{align}

Using the $s^*$ in (\ref{eq:effort}), we can indeed determine how the optimal payment plan varies with the accuracy of QoS measures for each resource and the operation cost coefficient of each resource. For example, when two resources are technologically substitution $c_{ij}>0$, if the measurability of resource $i$ worsens, that is, $\sigma_i^2$ increases, then, as is intuitive, $s_j^*$ goes up, but $s_i^*$ goes down. Thus, there is a measurement complementarity between the $s_i^*$ and $s_j^*$ in the presence of technologically substitutes problems \cite{Bolton2004}.

%\vspace{-0.3cm}
%%%%%%%%%%%%%%%%%%%%%%%%%%%%%%%%%%%%%%%%%%%%%%%%%%%%%%%%%%%%%%%
%%%%%%%%%%%%%%%%%%%%%%%%%%%%%%%%%%%%%%%%%%%%%%%%%%%%%%%%%%%%%%%
\section{Simulation Results and Analysis}\label{sec:Simulations}
%%%%%%%%%%%%%%%%%%%%%%%%%%%%%%%%%%%%%%%%%%%%%%%%%%%%%%%%%%%%%%%
%%%%%%%%%%%%%%%%%%%%%%%%%%%%%%%%%%%%%%%%%%%%%%%%%%%%%%%%%%%%%%%
%\vspace{-0.3cm}
In this section, we will first give a detailed analysis of how the NO's utility changes by varying the parameters such as the operation cost coefficients and measurement error covariance. Meanwhile, we will conduct a comparison of the NO's utility among different payment plans. To set up the simulation, we assume that, the reservation payment of the FN $\overline{w}=0$ when not cooperating in the fog computing ($a=0$). The reason we do not consider the FN's utility is that, from the optimal payment plan we have derived, no matter how those parameters change, the FN's utility will remain the same. The optimal payment plan will bring FN the utility the same as the reservation utility $-e^{-\eta\overline{w}}$, which in our case is $-1$ as we set $\overline{w}=0$.

%\vspace{-0.3cm}
In the previous section, we have solved the optimal payment plan when the measurement errors are stochastic dependent and the resource types are technologically dependent. As this multi-dimension case is the most general case in reality, we name this mechanism by \emph{General}. For comparison, we propose $5$ more payment plans. The fist one is the optimal payment plan when the measurement error and resource type are independent, and thus we name it by \emph{Independent}. The second payment plan is called \emph{Single Bonus} that is the payment plan obtained in the one-dimension case. In this one-dimension case, we can regard the NO payments FN on the QoS of a single one resource type. For example, measuring the data processing latency by only taking the CPU speed into account. The third and forth ones are special cases of the \emph{General}: one is stochastic independent but technologically dependent, the other one is technologically independent but stochastic dependent, and are named by \emph{Stochastic Independent} and \emph{Technologically Independent}, respectively. The last one is called \emph{Opening Reward}, which is the payment plan only contains a fixed salary $t$. We can regard this mechanism as the NO offers the FN a one time payment at subscription. But this \emph{Opening Reward} mechanism does not care about FN's future service quality.

In Fig. \ref{fig:costP}, we compare the NO's utility from the six payment plans as we vary the resource-specific operation cost coefficient $c_{ii}$. From the simulation results we see that, as the cost coefficient $c_{ii}$ increases, the NO's utility is decreasing in contrast. The reason for this phenomenon is that larger cost coefficient $c_{ii}$ means more operation cost when providing such an resource. Therefore, the FN is less likely to contribute in the fog computing. With less computing resources provided by the FN, the QoS will decrease and the NO's utility will certainly decrease. In addition, from Fig. \ref{fig:costP}, we see that the NO obtains the largest utility in the \emph{Opening Reward} case. Followed by the \emph{Independent}, \emph{Stochastic Independent}, and \emph{Technologically Independent}, the \emph{General} case proposed by us brings the fifth highest utility to the NO, while the \emph{Single Bonus} gives the least utility.

In Fig. \ref{fig:riskdgreeP}, we analyze the impact of FN's risk averse degree $\eta$ on the NO's utility. As the NO's utility $V=a-t$ in the \emph{Opening Reward} is independent of the risk averse degree $\eta$, we cannot see any change in the NO's utility. For the other five payment plans, we see that the NO's utility is decreasing as the FN's risk averse degree $\eta$ increases. This result is intuitive as a larger $\eta$ means the FN becomes more conservative and sensitive to risk, thus less likely to open access to end users. With less resources obtained from the FN, the NO's utility will certainly decrease. From Fig. \ref{fig:riskdgreeP} we also obtains the similar ranking of the NO's utility as in the previous figure: the \emph{Independent} case brings higher utility than the \emph{Stochastic Independent}, \emph{Technologically Independent}, and \emph{General} one, and the \emph{Single Bonus} one brings the smallest utility for the NO.

In Fig. \ref{fig:varianceP}, we increase the variance $\sigma_{i}^2$ to see how the NO's utility varies. Similar to the previous case, the NO's utility $V=a-t$ in the \emph{Opening Reward} is independent of the covariance matrix. Thus, we cannot see any change of the NO's utility. For the other payment plans, the NO's utility is decreasing with the variance, which is in accordance with our conclusion in the previous section. The variance $\sigma_{i}^2$ of measurement error denotes the relationship between the resources provided by the FN and the QoS observed by the NO. As $\sigma_{i}^2$ increases, it indicates a weaker relationship between resource quantity and the expected QoS achieved. As a result, the FN is likely to provide less amount of resource with increases in uncertainty, and thus a lower cost of cooperation. With the decrease of computing resources, the QoS is lowered and the NO's utility will certainly decrease. From Fig. \ref{fig:varianceP} we also obtain the similar ranking of the NO's utility as in the previous figure: the \emph{Independent} case brings higher utility than \emph{Stochastic Independent}, followed by \emph{Technologically Independent} and \emph{General} one, the \emph{Single Bonus} one brings the lowest utility for the NO.

The reason for the quality ranking of the six payment plans in Fig. \ref{fig:costP}, Fig. \ref{fig:riskdgreeP}, and Fig. \ref{fig:varianceP} is as follows. The \emph{Independent} payment plan is the ideal case of the \emph{General} multi-dimension case. As less measurement cost is occurred when predicting the QoS and less operation cost is encountered due to technology substitution, a higher utility is obtained than the other payment plans. The \emph{Stochastic Independent} and \emph{Technologically Independent} are partial independent cases of the \emph{General} multi-dimension one, thus, the NO's utility lies between the \emph{Independent} and \emph{General}. But as we have assigned larger values for the covariance matrix of the the measurement error than the operation cost coefficient matrix, more computing resources will be provided in the \emph{Stochastic Independent} than in the \emph{Technologically Independent}. Therefore, the NO's utility is higher in the \emph{Stochastic Independent} than in the \emph{Technologically Independent} case, while the \emph{Single Bonus} only payment FN with only one type of resource evaluation. As a result, the FNs have less incentive to provide computing resources. In return, less utility is obtained by the NO. For the result of the \emph{Opening Reward} case, it seems unreasonable at the first sight, as it brings the NO the highest utility than the other payment plans. While we notice that \emph{Opening Payment} is a ``once-for-all'' deal which does not provide continuous incentives for the FNs, i.e., after the FN has finished the computing task and receive the payment, it is likely to stop cooperating in the future.

%\vspace{-0.5cm}
%%%%%%%%%%%%%%%%%%%%%%%%%%%%%%%%%%%%%%%%%%%%%%%%%%%%%%%%%%%%%%%
%%%%%%%%%%%%%%%%%%%%%%%%%%%%%%%%%%%%%%%%%%%%%%%%%%%%%%%%%%%%%%%
\section{Conclusions}\label{sec:Conclusion}
%%%%%%%%%%%%%%%%%%%%%%%%%%%%%%%%%%%%%%%%%%%%%%%%%%%%%%%%%%%%%%%
%%%%%%%%%%%%%%%%%%%%%%%%%%%%%%%%%%%%%%%%%%%%%%%%%%%%%%%%%%%%%%%
%\vspace{-0.5cm}
In this paper, we have investigated the problem of maximizing NO's revenue by efficiently allocating FNs' computation resources in fog computing. The optimal payment plan is solved by paying the rent of FN's computing resources from a multi-dimension evaluation while ensuring the FN's cooperation. Furthermore, we use the numerical results to analyze the optimal payment plan by varying different parameters. In addition, we compare the NOs' utility under the six different payment plans, and show that the NO's utility deteriorates with large operation cost coefficient, higher risk aversion of FNs, and large measurement error variance.
%\vspace{-0.3cm}

%10\end{spacing}
\vspace{-0.2cm}
%\vspace{-0.5cm}
\bibliographystyle{IEEEtran}
\bibliography{./ICCS}

\end{document}